\documentclass[prl,twocolumn,superscriptaddress,preprintnumbers,a4paper,amsmath,amssymb,showpacs,floatfix]{revtex4}
\usepackage{graphicx}
\usepackage{bm}

\newcommand{\hb}{$H\|b$}

\newcommand{\Tn}{$T_{N}$}
\newcommand{\TN}{$T_{N}$}
\newcommand{\Tc}{$T_{c}$}
\newcommand{\Ts}{$T_{S}$}

\newcommand{\degg}{$^{\circ}$}
\newcommand{\pa}{$P\|a$}

\newcommand{\pc}{$P\|c$}

\newcommand{\Ps}{$\mathbf{P}_{s}$}

\newcommand{\mb}{$\mu_{B}$}
\newcommand{\mbmn}{$\mu_{B}/Mn$}
\newcommand{\bs}{$\mathbf{b}^{*}$}
\newcommand{\sxs}{$S_{i}\times S_{i+1}$}
\newcommand{\tmo}{TbMnO$_{3}$}

\begin{document}


\title{Magnetic field induced flop of cycloidal spin order in multiferroic \tmo: The magnetic structure of the \pa\ phase}

\author{N.~Aliouane}
\affiliation{Helmholtz-Zentrum Berlin f\"{u}r Materialen und Energy, Glienicker Str.~100, D-14109 Berlin, Germany}
\affiliation{Institute For Energy,  P.O. Box 40, NO-2027 Kjeller, Norway}

\author{K.~Schmalzl}
\affiliation{Institut f\"ur Festk\"orperforschung, Forschungszentrum J\"ulich GmbH,
JCNS at ILL, 38042 Grenoble Cedex 9, France}

\author{D.~Senff}
\affiliation{II. Physikalisches Institut, Universit\"at zu K\"oln,
Z\"ulpicher Str. 77, D-50937 K\"oln, Germany}

\author{A.~Maljuk}
\affiliation{Helmholtz-Zentrum Berlin f\"{u}r Materialen und Energy, Glienicker Str.~100, D-14109 Berlin, Germany}

\author{K.~Proke\v{s}}
\affiliation{Helmholtz-Zentrum Berlin f\"{u}r Materialen und Energy, Glienicker Str.~100, D-14109 Berlin, Germany}

\author{M. Braden}
\affiliation{II. Physikalisches Institut, Universit\"at zu K\"oln,
Z\"ulpicher Str. 77, D-50937 K\"oln, Germany}

\author{D.~N.~Argyriou}
\email{argyriou@helmholtz-berlin.de}
\affiliation{Helmholtz-Zentrum Berlin f\"{u}r Materialen und Energy, Glienicker Str.~100, D-14109 Berlin, Germany}

\date{\today}
\begin{abstract}
Using in-field single crystal neutron diffraction we have
determined the magnetic structure of \tmo\ in the high field \pa\
phase. We unambiguously establish that the ferroelectric
polarization arises from a cycloidal Mn spins ordering, with spins
rotating in the $ab$ plane. Our results demonstrate directly that
the flop of the ferroelectric polarization in \tmo\ with applied
magnetic field is caused from the flop of the Mn cycloidal plane.

\end{abstract}

\maketitle

\begin{figure}[htb!]
\includegraphics[scale=0.42]{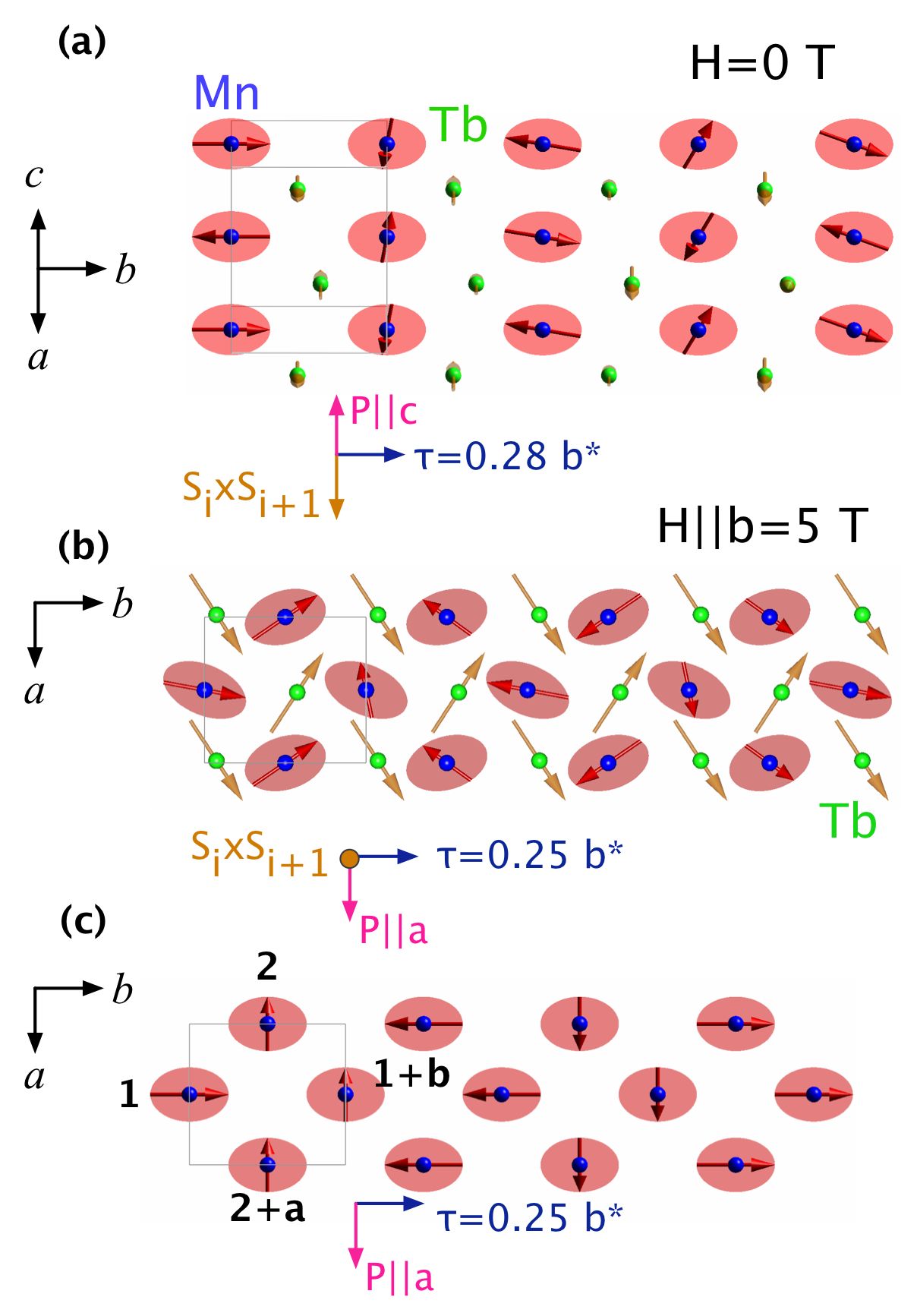}
\caption{(Color online) (a,b) Illustrations of the two cycloidal
Mn magnetic structures proposed to cause ferroelectricity in \tmo.
In both cases the magnetic propagation vector $\tau$ is parallel
to the $b-$axis. In zero field (panel (a)) cycloidal order, Mn
spins rotate around the $a$-axis (\sxs) and rotate wholly within
the $bc$-plane. The ferroelectric polarization via the
antisymmetric DM interaction is produced along the $c$-axis. The
high-field magnetic structure determined in this work is shown in
panel (b), it yields a \pa\ ferroelectric polarization that arises
from an $ab$ cycloid where Mn spins rotate around the $c-$axis. In
both panels we also show the magnetic ordering of Tb-spins.  In
the zero field case (a), the magnetic propagation vectors of Tb
and Mn-spins are clamped and Tb-spins point along the $a$-axis
forming a SDW \cite{Cheong:2007dw}. The canted antiferromagnetic ordering of Tb-spins
for \hb=5T determined in this work is shown in panel (b). (c)  Here we depict an anharmonic $ab$-spiral where the phase difference between spins 1 and 2 is $\omega=\pi/2$. Is such a case the angle between spins 1 and 2 is different from that between 2 and 1+b yielding an alternating scalar product along the $b-$axis. Amplitudes of Tb and Mn spins are not to scale.
}
\label{cycloids}
\end{figure}

The antisymmetric Dzyaloshinski-Moriya (DM)
interaction\cite{Dzyaloshinsky:1958km,moriya} between two spins,
$S_{i}$, $S_{i+1}$ separated by $r_{i,i+1}$, provides for a
natural coupling between magnetism and ferroelectricity with the
spontaneous ferroelectric polarization given by
$\mathbf{P_{s}}\sim \mathbf{r}_{i,i+1}\times (\mathbf{S}_{i}\times
\mathbf{S}_{i+1})$
\cite{mostovoy:067601,katsura:057205,Cheong:2007dw}. This
mechanism generates ferroelectricity in a wide  variety of magnets
such as $R$MnO$_{3}$~ perovskites with $R$=Gd, Dy, and
Tb,\cite{kimura,goto}, spinel chromate
CoCr$_2$O$_4$,\cite{Yamasaki:2006zf} spin-chain cuprate
LiCu$_2$O$_2$,\cite{Park:2007bh} and huebnerite
MnWO$_4$\cite{Heyer:2006oj,Taniguchi:2006eq}. In the $R$MnO$_{3}$
manganites the DM interaction results in cycloidal order of
Mn-spins giving a spontaneous ferroelectric polarization along the
$c$-axis (\pc) (Fig.~\ref{cycloids}(a)). The application of
magnetic field results in the flop of the polarization from the
$c$- to the $a$-axis and highlights a novel control of one ferroic
property by an other\cite{kimura}. It has been assumed that this
change in the direction of the polarization reflects the flop of
the Mn spin cycloid, implying that the antisymmetric DM
interaction continues to be responsible for the polarization
(Fig.~\ref{cycloids}(b))\cite{mostovoy:067601,senff:174419}.
However, in this high field \pa-phase, the commensurate magnetic
wave vector for Mn is also compatible with other magneto-electric
mechanisms such as exchange
striction\cite{Sergienko:2006yo,Sergienko:2006qc,aliouane:020102}.
In this letter we present a determination of the magnetic
structure of \tmo\ in a high magnetic field in the commensurate
\pa\ phase. We find that the magnetic structure of Mn-spins is
characterized by an $ab$-cycloid that accurately describes the
direction of the observed ferroelectric polarization via the  DM
interaction. Our finding validates the model that the polarization
flops found in the perovskite manganites result from the flop of
the Mn-spin cycloid. 

The manganite \tmo\ crystallizes in the orthorhombic perovskites structure P$bnm$.
On cooling, below \Tn=41K Mn spins order incommensurately point along the magnetic
wave vector $\tau\sim0.275$\bs \cite{Cheong:2007dw,kimura}. On further cooling below
\Ts=28K, a $c$-axis component of the Mn moment orders with a phase shift of $\pi/2$
with respect to the $b$-component so as to form a cycloidal  structure where Mn-spins
rotate within the $bc$ plane and around the $a$-axis as shown in Fig.~\ref{cycloids}(a)\cite{Cheong:2007dw}.
The axis of spin rotation defines the DM interaction, $\mathbf{S}_{i}\times \mathbf{S}_{i+1}$,
while the distance $\mathbf{r}_{i,i+1}$ is parallel to the modulation vector $\tau$.
For this type of spin order, inversion symmetry is broken yielding for $R$=Tb and Dy, 
a ferroelectric polarization along the $c$-axis as indeed is observed ($\mathbf{P_{s}}\|\mathbf{c}\sim\mathbf{a}\times\mathbf{b}$)\cite{Cheong:2007dw,kimura}. 

It is tempting to assume that the flop in the ferroelectric
polarization arises from a flop in the Mn-spin spiral, but so far
there is no experimental prove for this. Furthermore, the field
dependence of the Mn-spin spiral is difficult to analyse as
single-ion anisotropy terms of Mn and $R$ need to be taken into
consideration as well as the $R$-Mn interaction. The fact that in
this \pa\ phase Mn-spins order commensurately with
$\tau=\frac{1}{4}$\bs\cite{aliouane:020102,Arima:2005ir} render
the cycloidal flop model even more complex. For such a
commensurate spin structure it has been proposed that spin
frustration and super-exchange induce lattice distortions that
break inversion symmetry thereby generating the observed direction
of the polarization at high field
\cite{aliouane:020102,Sergienko:2006yo,Sergienko:2006qc}. This
exchange striction model can be applied only to a commensurate
order, ruling out its validity for
$R$=Dy\cite{Strempfer:2007p172}. The observation that infrared
electromagnon signals are always polarized along the $a-$axis
irrespective if \Ps\ is parallel to the $a$- or $c$-axis further
adds to the debate of the high field magnetic phases in these
manganites\cite{Kida:2008p8562}. A cycloidal Mn magnetic ordering
that yields \pa\ in zero field has been observed for
Gd$_{0.7}$Tb$_{0.3}$MnO$_{3}$\cite{yamasaki:097204}, however,
there is no direct evidence that the magnetic field flops the
Mn-cycloid to yield a \pa\ polarization. It is therefore pressing
to establish an accurate model of the high field magnetic
structure of these multiferroic manganites.

\begin{figure}
\includegraphics[scale=0.25]{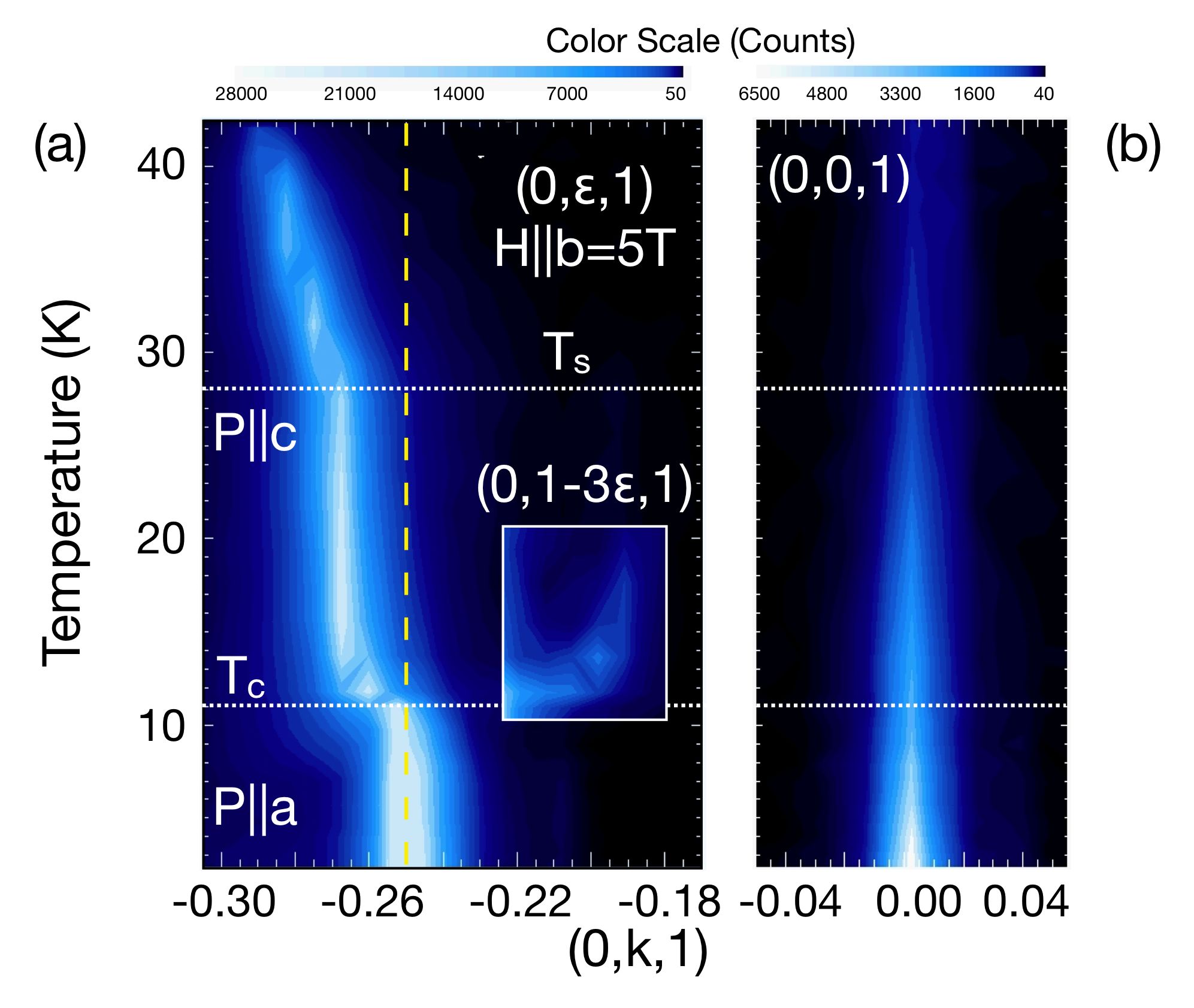}
\caption{(Color online) Single crystal neutron diffraction
measurements from the  E4 diffractometer. Here we show scans along
(0,$k$,1) in reciprocal space as a function of temperature
measured in a 5T field applied along the $b$-axis.  The intensity
of these scans is plotted in color coding with corresponding
scales above each panel. (a) Portion of the data showing the
temperature dependence of the A-mode reflection (0,$\epsilon$,1).
Here the wave number $\epsilon$ varies from 0.282 at \TN\ to 0.266
at 15 K before it locks discontinuously to the commensurate value
of $\frac{1}{4}$ below 11K. The weak third harmonic reflection was
also observed in these scans to disappear at this transition.
These data are shown in an enhanced color scale on the same panel
and in the same location in the $T-Q$ map. (b) Temperature
evolution of the P$bnm$ forbidden reflection (0,0,1).}
\label{data}
\end{figure}

\tmo\ single crystals were obtained by re-crystalizing a ceramic
rod under Ar atmosphere using an optical floating-zone furnace.
The field and temperature dependence of the magnetic propagation
wave vector was measured on the E4 double axis neutron
diffractometer at the BENSC facility of the Helmholtz Zentrum
Berlin using a neutron wave length of $\lambda$=2.45\AA\ and a
$\lambda/2$ filter placed in the incident beam. Here magnetic
field was applied horizontally along the $b-$axis using the HM1
superconducting cryo-magnet. Due to the limited view of the sample
in this magnet, integrated intensities of Bragg reflections could
only be measured within certain regions of the $0kl$ lattice
plane, a situation that prohibits accurate analysis of the
magnetic structure.  To overcome this problem we conducted
measurements using the D23 neutron single-crystal diffractometer
installed at the Institut Laue-Langevin (ILL) with
$\lambda=1.281$\AA,  that is equipped with a lifting detector that
allows measurements of Bragg reflections above and below the
scattering plane. Magnetic field was applied using the 6T vertical
field superconducting magnet with an asymmetric vertical opening
angle of -5/+10$^{\circ}$. For these measurements a single crystal
of \tmo\ was cut into a parallelepiped with dimension
3.0x2.9x3.7mm with each face perpendicular to one crystallographic
direction. The crystal was oriented with the $b$-axis parallel to
the field. This geometry allowed us to measure reflections with
$k$ from -1.8 to 0.25.  In total 138 independent nuclear Bragg
reflections consisting of 310 individual reflections were
collected at $H=0T$ and T=8.5K within a range of
($0.07\leqslant\sin(\theta)/\lambda\leqq0.69)$. At 8.5K the
magnetic field was applied along the $b$-axis to 5T. In these
conditions 140 commensurate ($\tau=0$) independent Bragg
reflections consisting of 160 individual reflections were measured
along with 64 independent reflections with $\tau=\frac{1}{4}$\bs.
Analytical absorption corrections for each reflection were made by
the Xtal suite of program, while analysis of the magnetic
intensities was performed with the Fullprof code.

In Fig.~\ref{data}(a) we show  measurements of the
(0,$\epsilon$,1) reflection as a function of temperature with
\hb=5T measured on the E4 diffractometer.  The data shows that
Mn-spin ordering is first observed at \TN=41K with the wave number
$\epsilon$ decreasing on cooling.  At \Tc=11K, $\epsilon$ changes
discontinuously  to yield a commensurate wave vector of
$\tau=\frac{1}{4}$\bs\cite{Senff:2007p176,aliouane:020102,Arima:2005ir}.
Published polarization data shows that a \pc\ state develops below
\Ts\ while the polarization flops to \pa\ at \Tc, coinciding with
the transition to the commensurate magnetic phase\cite{kimura3}.
In the same data we observe that the third harmonic reflection
(0,1-3$\epsilon$,1) rapidly changes its position and disappears
also at \Tc (Fig.~\ref{data}(a)). On cooling below \Tn\ we find an
enhancement in the intensity of several nuclear reflections and
the appearance of forbidden reflections such as the (0,0,1) shown
in Fig.~\ref{data}(b). As we discuss below these effects arise
from the ordering of Tb-spins with magnetic propagation vector
$\tau^{Tb}=0$.

\begin{figure}
\includegraphics[scale=0.7]{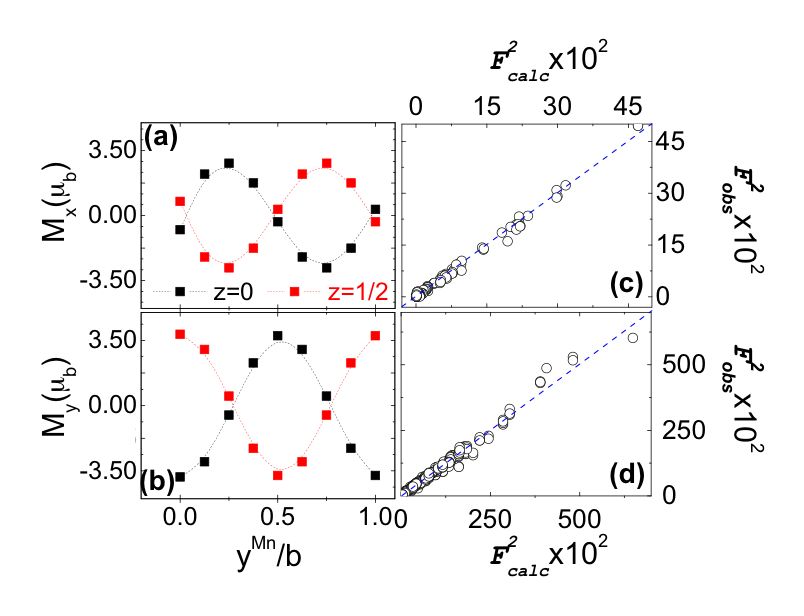}
\caption{ (Color online) Results of refinements of the magnetic
structure at 8.5K and \hb=5T. In panels (a) and (b) we show the
variation of the $M_{x}$ and $M_{y}$ components of the Mn-spins
for the ions located at $z$=0 (black) and $z=\frac{1}{2}$ (red)
along the magnetic propagation vector. Note that the components
along $a$ and $b$ are out of phase so as to yield a cycloidal
ordering shown in Fig.~\ref{cycloids}.  In panels (c) an (d) we
show the comparison between observed and calculated magnetic
structure factors ($F^{2}$) for the analysis of the Mn
$\tau=\frac{1}{4}$ reflections and nuclear plus Tb-magnetic
reflections respectively. Since the Tb magnetic propagation vector
is $\tau=$0 the nuclear and magnetic reflections are not separated
in reciprocal space as in the case for the Mn-ions.}
\label{mnspin}
\end{figure}

To determine  the magnetic  structure of \tmo\ in the \pa\ state
we utilized the D23 diffractometer. Here the sample was cooled in
zero field to 8.5K and then a \hb=5T was applied so as to enter
the commensurate \pa\ phase that was confirmed by measurements of
the magnetic wave vector. At this field and temperature we find
that the most intense magnetic reflections with wave number
$\epsilon=\frac{1}{4}$, have extinction condition $h+k=$even,
$l$=odd (A-mode)\footnote{$h,k,l$ are Miller indices at the center
of the primitive Brillouin zone}, while reflections with
$h+k=$odd, $l=$odd (G-mode) were considerably weaker. For space
group P$bnm$ and wave vector $\tau\leqslant\frac{1}{4}$\bs\ there
are four irreducible representations (irreps) $\Gamma$ of the
magnetic symmetry for the Mn-ion \cite{bertaut,Brinks:2001jo}. The
A mode reflections are contained only in irreps $\Gamma_{1}$,
$\Gamma_{2}$ and $\Gamma_{3}$. Analysis of the data using only
A-mode reflections and a single irrep did not result in a
satisfactory fit to the data.  This lead us to consider
combinations of representations.  Of the possible combinations
only a model using $\Gamma_{1}\otimes\Gamma_{3}$ produced a
satisfactory result with $R(F^{2})$=7.9\% and
$R_{w}(F^{2})$=8.5\%.  This coupled  irrep has the form of
$(A_{x}, G_{y}, C_{z}) \otimes (G_{x}, A_{y}, F_{z} )$. Our
measurements showed that F- and C- modes were at the very limit of
detection indicating that the Mn moment is essentially contained
within the $ab$-plane.  Reflections from these modes were not
included in the final refinements. \footnote{Reflections condition
for the C-mode are $h+k=$even, $l=$even  and for the F-mode $h+k=$odd, $l=$even.} Our analysis yielded a magnetic structure for Mn
given by the moment in $\mu _B$:
$m^{Mn}=\Gamma_{1}(2.83(12),0.51(4),0)\otimes\Gamma_{3}(0.55(4),3.79(7),0)$.
The phase between the two irreps is also a variable parameter
found to be $0.474(12)\pi$ close to $\pi \over 2$ expected for a
cycloid. \footnote{Our sensitivity to the phase difference was
tested by fixing it to a value of $\pi/2$ which lead to
statistically significantly higher R-factors of $R(F^{2})$=8.4\%
and $R_{w}(F^{2})$=9\%.}  The values of the Mn-spins in this
commensurate cycloidal structure along the $a-$ and $b-$axis are
given in Fig.~\ref{mnspin} for the $z$=0 and $z=\frac{1}{2}$
layers.

The magnetic structure indicated by this model is dominated by the
A modes of these two irreps ($A_{x}$ and $A_{y}$) producing an
elliptical cycloid, shown in Fig.~\ref{cycloids}(b). Here the Mn
cycloid is contained within the $ab$-plane and Mn-spins rotate
around the $c-$axis, consistent with the direction of the
ferroelectric polarization along the $a-$axis
($\mathbf{P}\sim\mathbf{b}\times\mathbf{c}$). The anisotropy of
the Mn cycloid in both low field $bc$ and high field $ab$
configurations appears to be very similar.  In both cases the
$A_{y}$ mode possesses the higher moment (3.9 and 3.79 \mbmn\
respectively) while the components orthogonal to this mode are
smaller and of the same magnitude (2.8\mbmn).  This result shows
that the flop of the cycloidal plane does not effect the
anisotropy of the Mn ordering itself. The deviation of the phase
shift between the two irreps away from the ideal value produces
an angle between spins of $85^{o}$. Despite this, $\mathbf{ S_{i}}\times\mathbf{
S_{i+1}}$ remains parallel to the $c-$axis and should not
influence the magnitude of \Ps\ significantly. Finally for the Mn
ordering we find that the $G_{x}$ and $G_{y}$ modes are active
with amplitudes of $\sim$0.5\mbmn\ and its net effect on the over
all cycloidal order that produces a polarization along the
$a-$axis is also relatively small.

We now turn our attention to the Tb-spin ordering. We find that in
the \pa\ phase Tb-spins order commensurately with the underlying
primitive lattice (i.e. $\tau^{Tb}=0$).  Analysis of the
commensurate P$bnm$ reflections clearly showed additional
intensity that can be modelled by the Tb magnetic order. The best
fit to the measured data was obtained for a ferromagnetic
alignment of Tb-spins along the $b$-axis and antiferromagnetic
coupling between nearest-neighbor
Tb-spins along the [110] direction (Fig.~\ref{cycloids}(b)). Analysis of this structure yields a total Tb moment of 7.24(7)\mb, with an antiferromagnetic component of 6.07(9)\mb\ along the $a$-axis and a ferromagnetic component of 3.92(6)\mb\ along the $b$-axis. The ferromagnetic ordering along the $b$-axis is indeed evident in magnetization measurements under similar conditions\cite{kimura3}. 

The work we present here unambiguously proves that the
commensurate \pa\ phase in \tmo\ coincides with an $ab$ Mn spin
cycloid for \hb=5T. The antisymmetric DM interaction in this case
does yield a ferroelectric polarization along the $a-$axis as
indeed is observed
($\mathbf{P_{s}}\|\mathbf{a}=\mathbf{\tau}\times(\mathbf{S}_{i}\times
\mathbf{S}_{i+1})=\mathbf{c}\times\mathbf{b}$).
We may thus identify the inverse DM interaction as the main
mechanism for the magnetic-field induced flop of ferroelectric
polarization.

In a perfect cycloidal magnetic arrangement the exchange mechanism
proposed in references \cite{Sergienko:2006yo,Sergienko:2006qc}
does not yield any ferroelectric polarisation, as the scalar
product ($\mathbf{S}_{i}\cdot \mathbf{S}_{i+1}$) of neighboring spins is everywhere the same. This still
holds for the commensurate perfectly circular spiral. In the case
of an elliptical commensurate cycloid the exchange mechanism does
cause ferroelectric polarization as the scalar product varies
along the modulation. In our case with a modulation of four
orthorhombic lattice distances the mechanism of
\cite{Sergienko:2006yo,Sergienko:2006qc} may thus yield a finite
polarisation along the $a$-direction which, however, should still
be small due to the only minor deviation from a perfect
circular cycloid. The exchange mechanism may gain further
importance in the case of a very \emph{anharmonic} cycloid. In the
extreme anharmonic arrangement, where spins point either along the
$b$ or $a$ directions yielding the sequence along the b-direction shown in fig.~\ref{cycloids}(c):
left, up, right, down, there will be a very
effective exchange striction mechanism as the scalar product $\mathbf{S}_{i}\cdot \mathbf{S}_{i+1}$, between nearest neighboring Mn ions alternates along the $b$-axis (in the example of fig.~\ref{cycloids}(c) the alternating angles are 90 and 0 deg. along the $b$-direction). In the current analysis we have performed, the phase difference between rows of cycloids propagating along the $b$-axis (e.g. in Fig.~\ref{cycloids}(c), rows starting at ions 1 and 2) is fixed by symmetry as $\omega=\pi\mathbf{\tau}/2=45$\degg \cite{Brinks:2001jo}\footnote{Here $\omega$ should not be confused with the phase difference between irreps $\Gamma_{1}$ and $\Gamma_{3}$.}. A deviation from this value will yield an anharmonic spiral, however such as case can not be fully tested with the current data and we can not exclude a small degree of anharmonicity. In view of the current debate about the electromagnon it appears interesting to add that this exchange-striction mechanism will
always yield ferroelectric polarization along the $a$-direction.

In conclusion our work demonstrates that the flop in the
ferroelectric polarization observed in \tmo\ arises from the flop
of the Mn cycloidal plane from $bc$ to $ab$. The cycloidal
magnetic structure we establish here for the high field \pa\ phase
does not posses a dominant ferroelectric mechanism based on exchange striction.
The ordering of Tb-spins in this high field phase is that of a
canted antiferromagnet giving a significant ferromagnetic component
along the $b$-axis. 

We acknowledge the assistance of L.C. Chapon and Juan Rodriguez-Carvajal with Fp-studio and Fullprof respectively.  We thank M. Mostovoy and D. Khomskii for helpful discussions.


\end{document}